\documentclass{aa} 
\usepackage{graphicx}
\usepackage{txfonts}
\usepackage[colorlinks=true,citecolor=blue]{hyperref}
\usepackage{graphicx}
\usepackage{times}
\usepackage{natbib}
\usepackage{color}
\usepackage{multirow}
\usepackage{todonotes}
\usepackage{appendix}
\usepackage{array}
\usepackage{verbatim}

\begin{document} 

\title{On the interpretation of XRISM X-ray measurements of turbulence in the intracluster medium: a comparison with cosmological simulations}
\titlerunning{Turbulence in the intracluster medium: in X-rays and in simulations} 
\author{F. Vazza\inst{1,2} \fnmsep\thanks{\email{franco.vazza2@unibo.it}}
          \and
          G. Brunetti\inst{2} 
}
\institute{
   Dipartimento di Fisica e Astronomia, Università di Bologna, Via Gobetti 92/3, 40121 Bologna, Italy
         \and
            Istituto di Radio Astronomia, INAF, Via Gobetti 101, 40121 Bologna, Italy
        }

   \date{Received / Accepted}

\abstract    
{We investigate whether the properties of turbulent gas motions recently measured via X-ray spectroscopy in the Coma cluster of galaxies by XRISM are in tension with the turbulent picture established by current numerical cosmological simulations.
We use a high-resolution simulation of a Coma-like cluster and show that the simulation yields  velocity structure functions and X-ray line-widths that are compatible with those measured by the XRISM observations of Coma.
In particular, it has been previously suggested that a much steeper turbulence spectrum than the Kolmogorov would be needed to explain the XRISM observations under a homogeneous, cluster volume-filling turbulence model. Our results show that this tension is overcome thanks to the more complicated turbulent picture in cosmological simulations,  that indeed shows a patchy distribution of turbulent regions in galaxy clusters, with a spectrum that is generally consistent with a Kolmogorov power-law over a fairly wide range of scales.
More generally, our study highlights the fact that the interpretation of XRISM data of galaxy clusters depends on the turbulence model used and the importance of combining data and advanced simulations in the future steps.}


\maketitle

\begin{figure*}
\includegraphics[width=0.94\textwidth]{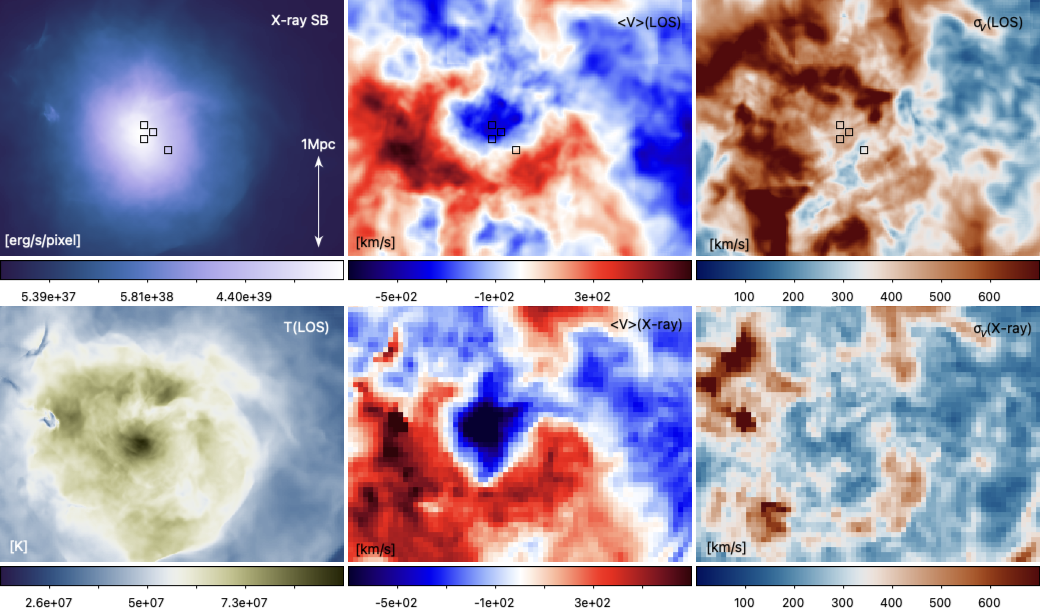}
\caption{Projected X-ray surface brightness in the [5-7] keV energy range  for our simulated cluster (top left), average X-ray weighted gas temperature (lower left),  average gas velocity along the line of sight, either using a volume-weighting procedure (top centre) or an X-ray emissivity weighting along the line of sight, within pixel of $90 \times 90 \rm ~kpc^2$ to mimic XRISM field of view for Coma (bottom centre). Gas velocity dispersion along the line of sight, again either using a volume-weighting procedure (top right) or an X-ray emissivity weighting along the line of sight, for $90 \times 90 \rm ~kpc^2$ pixels (bottom right). The black squares refer to the regions used to produce the line emission profiles given in Fig.\ref{fig:line}.}
 \label{fig:map0}
\end{figure*}

\section{Introduction}
\label{sec:intro}

Turbulence in the intracluster medium (ICM) is expected to  arise from the stirring associated with the mass growth of clusters
\citep[e.g.][]{bn99,su06}, and may provide pressure support in the atmosphere of clusters of galaxies \citep[e.g.][]{gb01,petrosian01, 2013ApJ...771..102F,2016ApJ...827..112B,2019A&A...621A..40E}.  Turbulence may also play a role for the origin of non-thermal components in the ICM, leading to the amplification of magnetic fields \citep[e.g.][]{do99,sch05, cho14,bm16} and to the re-acceleration of relativistic particles \citep[e.g.][]{2004MNRAS.350.1174B,cassano05,2015ApJ...800...60M,   2016MNRAS.458.2584B, beduzzi24,nishiwaki24}. 
The turbulence in clusters of galaxies has been since long studied with cosmological hydrodynamical simulations  \citep[e.g.][]{bn99,do05,in08,va11turbo,miniati14}, albeit its precise determination depends on the specific algorithms used to filter out bulk motions, shocks and other perturbations from the 3-dimensional velocity field \citep[e.g. see  discussion in][]{va12filter,2014ApJ...792...25N,va17turb,va18turb,2020MNRAS.495..864A,2021MNRAS.504..510V}.

The observed fluctuations of thermodynamic quantities derived from 
 X-rays and from the Sunyaev-Zeldovich effect in clusters of galaxies were interpreted as indications of moderate fluctuations induced by turbulence \citep[e.g.][]{sc04,2012MNRAS.421.1123C,ga13,2012MNRAS.422.2712Z,2016MNRAS.463..655K,2017ApJ...843L..29E,2024A&A...687A..58D,2024A&A...682A..45L}. 
Additionally, recent works also analysed the trails of ionized interstellar medium swept up behind 
"jellyfish" galaxies to trace turbulence  of the surrounding ICM, and inferred
the presence of $\sim 20-100 \rm ~km/s$ on $1-60 \rm ~kpc$ scales, with a distribution compatible with Kolmogorov turbulence \citep[][]{2023MNRAS.521.4785L,2024ApJ...977..219I}.
 
However,  the direct measurement of turbulence in the ICM has only since recently become possible, with spectroscopic X-ray observations of a few clusters. 
The Hitomi satellite managed to detect root-mean square velocities in the (fairly relaxed) Perseus cluster of $\sim 200 \rm ~km/s$ on $\leq 60 ~\rm kpc$ \citep[e.g.][]{hitomi,zuhone18}.
More recently, the XRISM collaboration reported first measurements of bulk velocity and of the velocity dispersion along a few lines of sight (LOS) crossing three galaxy clusters in the local Universe: A2029 \citep[][]{2025arXiv250506533X,2025ApJ...982L...5X}, Coma \citep[][]{Coma_XRISM}, Centaurus \citep[][]{2025Natur.638..365X}. In all cases, the reported values of LOS velocity dispersion (once converted into three-dimensional values, assuming isotropic velocity distributions) are in the $\sim 200-400 ~\rm km/s$ range at all radii, resulting in typical inferred kinetic to total pressure support of $\sim 2-3 \%$, 
naively resulting in a much smaller estimate than in previous simulations.
A key finding obtained by XRISM in the case of the Coma cluster (a massive perturbed system hosting multiple diffuse radio emissions, \citealt{bonafede21} for recent observations)  is that, while the slope of velocity structure function they derive from observations is poorly constrained, the combination of the velocity structure function and line broadening requires a turbulent spectrum that is much steeper than a Kolmogorov, if a homogeneous turbulent model is assumed \citep[see][for an updated discussion]{2025arXiv251021918E}. This is used to propose either a very large effective viscosity in the ICM inducing a large dissipation scale of turbulent motions, or a situation where large scale motions generated by recent mergers have not yet cascaded down to small scales \citep[][]{Coma_XRISM}.
If confirmed, these findings would start providing fundamental insights on the properties of turbulence in the ICM and on the role of collision-less effects. 
Indeed in a weakly-collisional plasma instabilities driven by large scale motions may govern the effective viscosity of the plasma and the fraction of the energy that is drained from large scales into small scale turbulence \citep[e.g.][]{2011MNRAS.410.2446K,bl11,2016PNAS..113.3950R,2018ApJ...863L..25S,2020JPlPh..86e9003S,2025A&A...694A..25K}.

Given the potential impact of these findings, in this article we attempt to explore how effective the current XRISM measurements really are in obtaining constraints on the turbulence in the ICM and whether they are really in tension with current cosmological simulations which are based on a purely collisional model of the ICM. The paper is structured as follows: in Sec.~\ref{sec:methods} we describe our simulation; in Sec.~\ref{sec:res} we give our results and in Sec.~\ref{sec:conclusions} we discuss the implications of this work for the interpretation of existing and future observations. 


\begin{figure}
\includegraphics[width=0.495\textwidth]{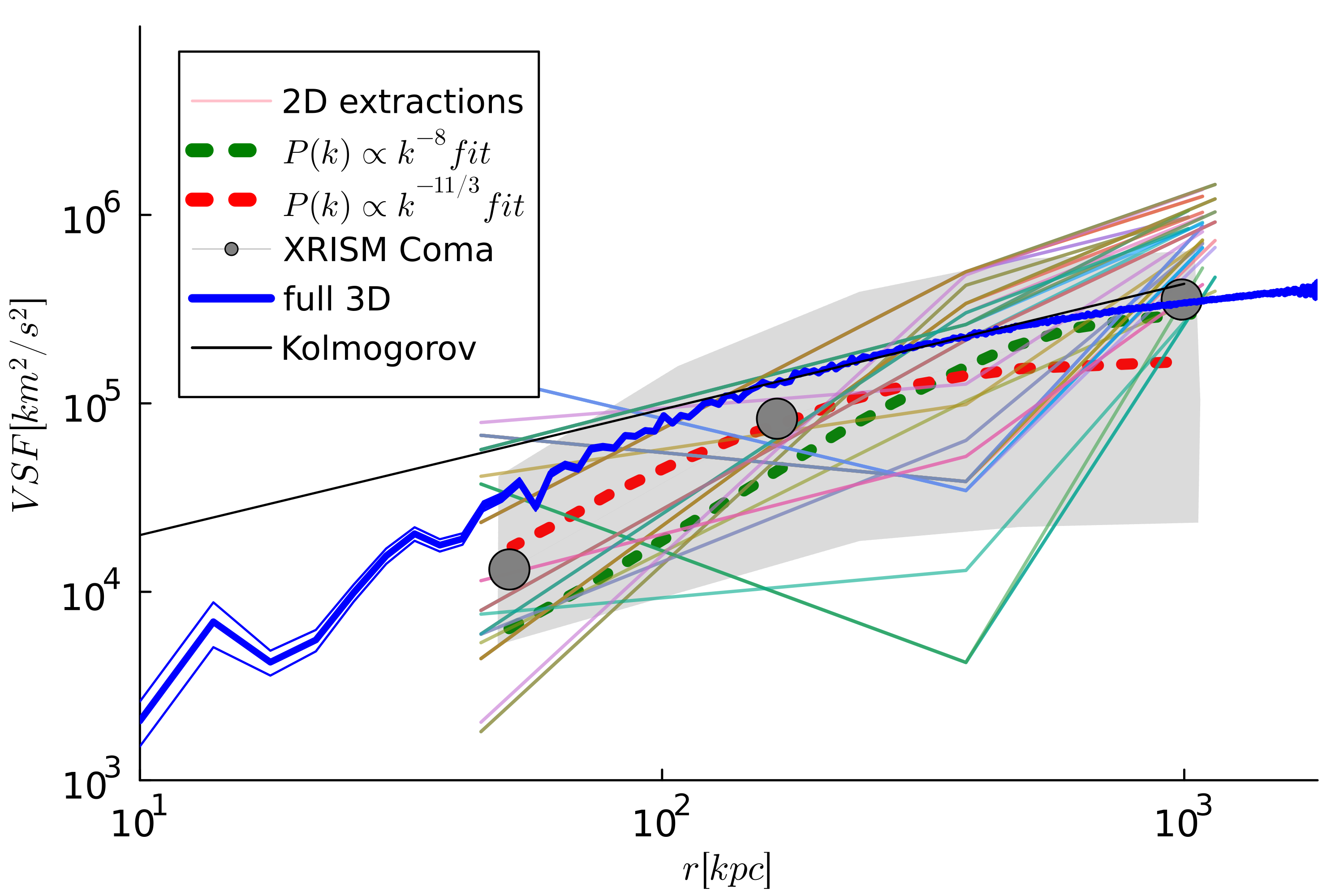}
\caption{Thin lines with different colours: simulated VSFs for 50 random  lines of sight, with area $90 \times 90 \rm ~kpc^2$, randomly extracted from our 2D velocity map and with  separations as in the XRISM observation of Coma. Blue lines: full 3-dimensional VSF for the velocity along the line of sight component, using $10^7$ cells in the  cluster volume. The grey points are the XRISM measurements for Coma, the additional dashed lines show the best-fit models derived \citet{Coma_XRISM} for a Kolmogorov spectrum (red) or a much steeper one (green), while the grey shaded area show the 68\% cosmic variance uncertainty and the measurement statistical errors in the same work.}
 \label{fig:VSF}
\end{figure}

\begin{figure}
\includegraphics[width=0.49\textwidth]{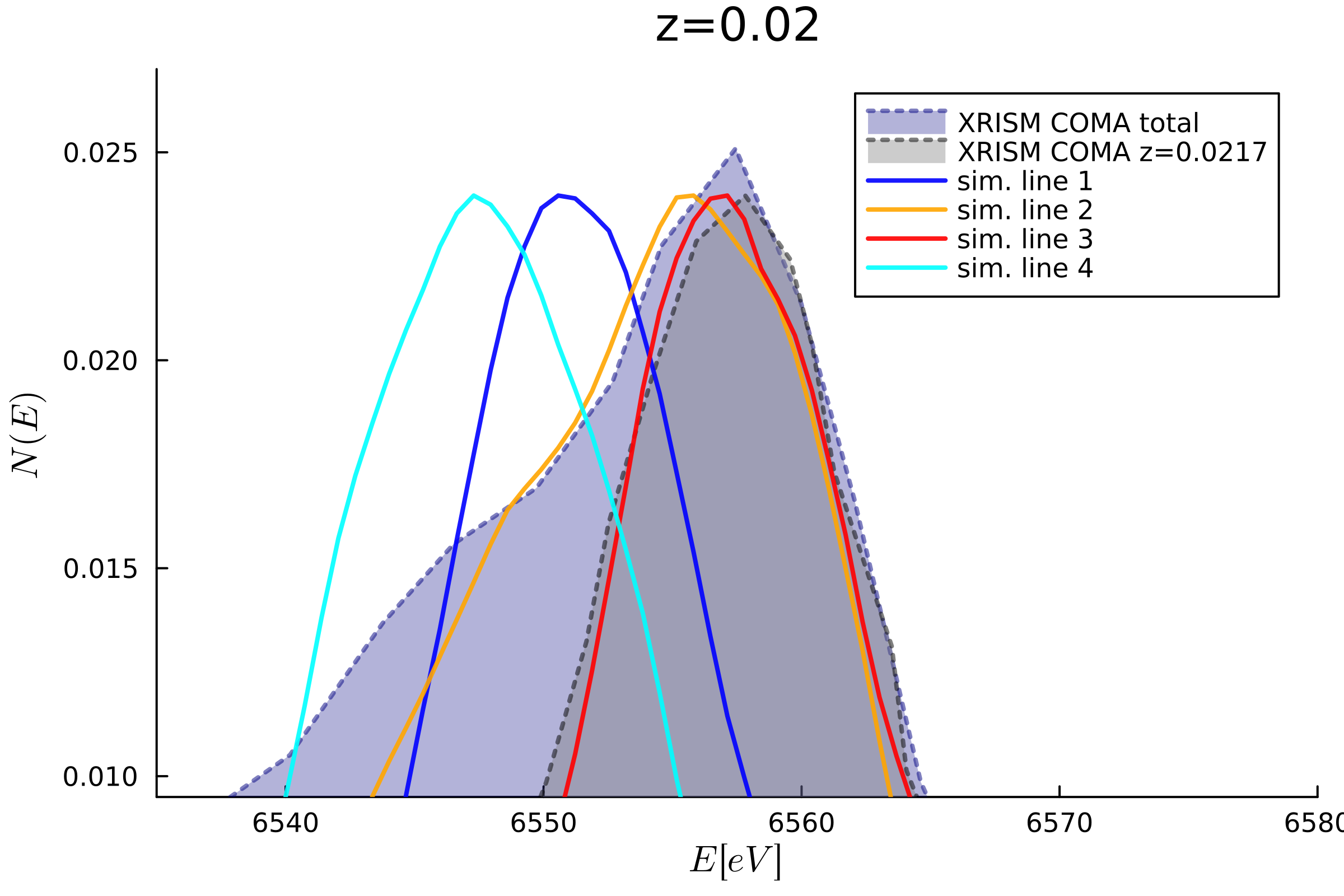}
\caption{Simulated line profiles for the four FOVs indicated in Fig.\ref{fig:map0} (color lines) compared with the 
reconstructed line models for the XRISM observation of the Coma cluster produced by \citet{Coma_XRISM}, given by the two shaded areas.} 
 \label{fig:line}
\end{figure}

\begin{figure}

\includegraphics[width=0.49\textwidth]{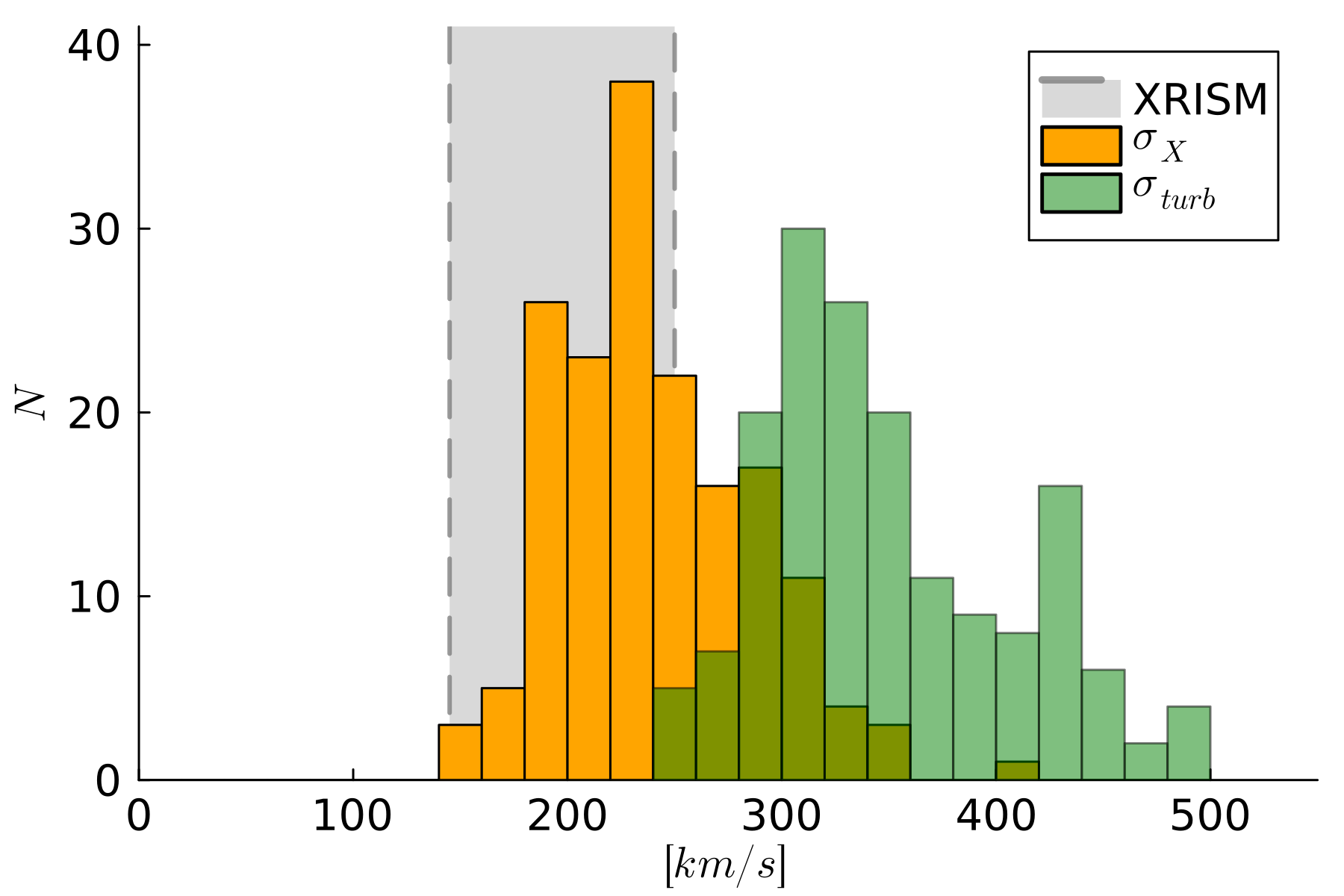}
\includegraphics[width=0.49\textwidth]{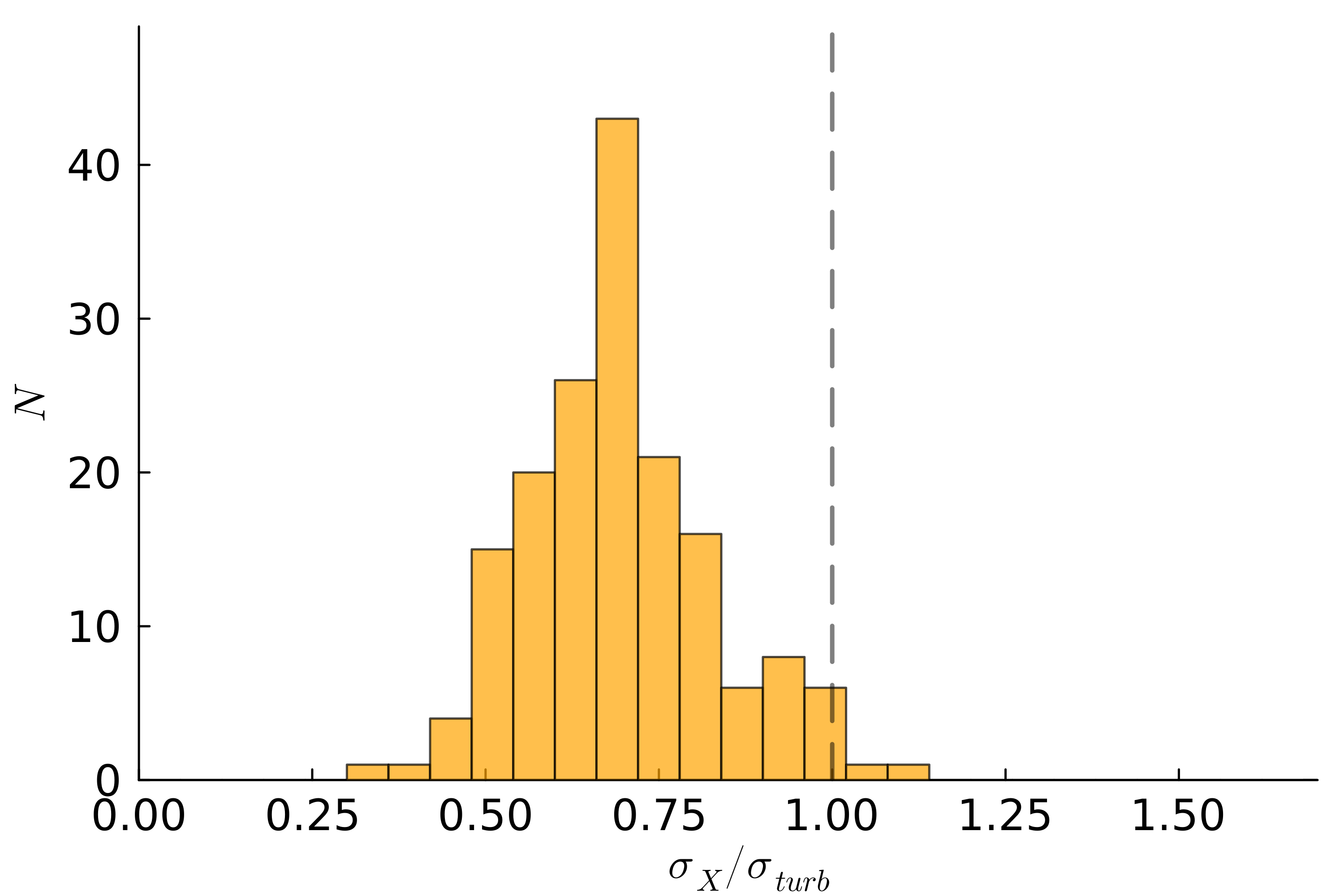}
\includegraphics[width=0.49\textwidth]{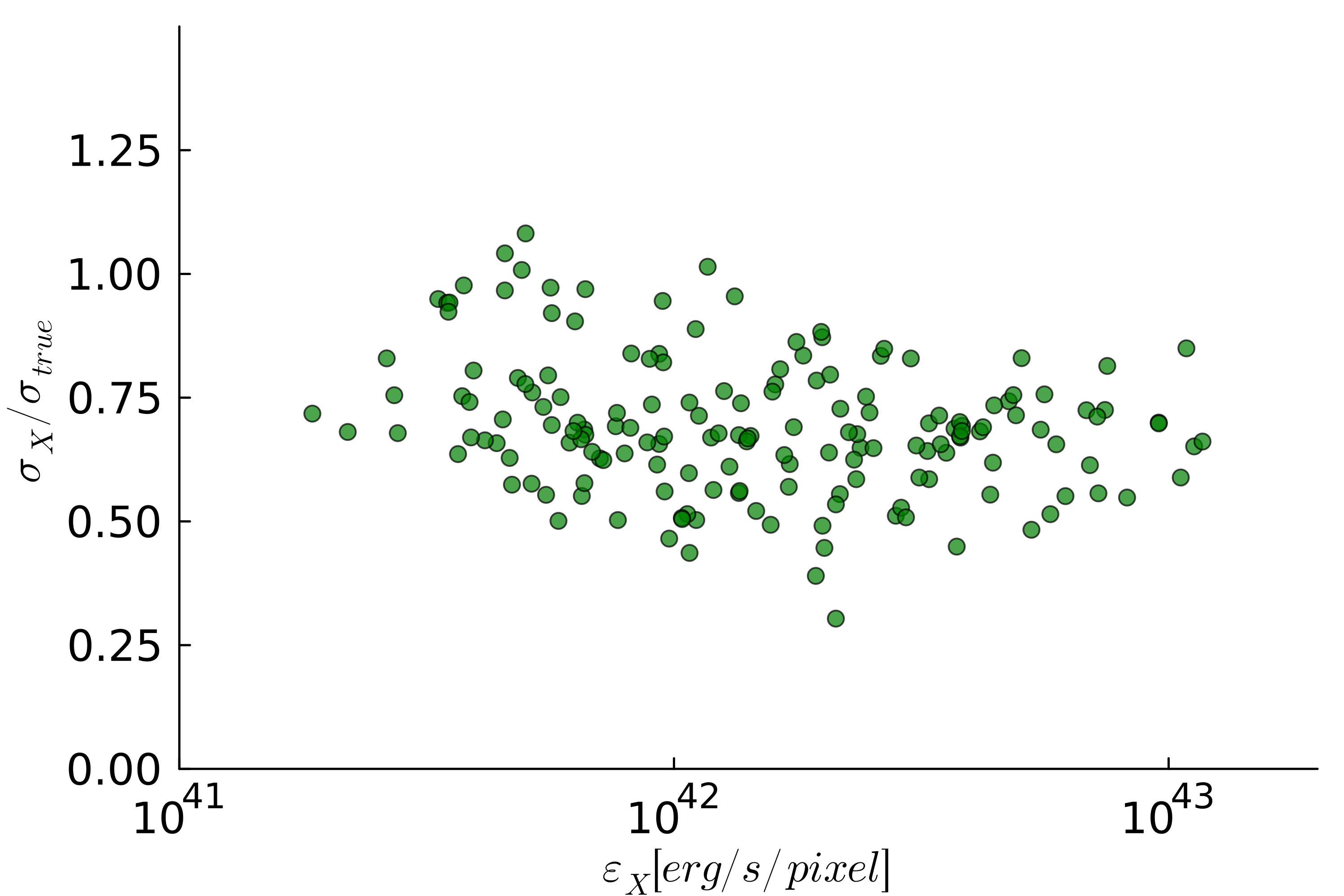}
\caption{Top panel: histograms of X-ray weighted velocity dispersion along the LOS and of volume weighted filtered turbulent velocity dispersion (after the application of our small-scale filtering with a $\Lambda=300 \rm ~kpc$ scale) for $90 \times 90 \rm ~kpc^2$ pixels maps. The additional vertical lines give the values inferred for the central FOVs observed by XRISM in the central region of Coma. Central panel: histograms for the ratio between the two estimates of the velocity dispersion above.  Bottom panel: ratio between the X-ray weighted velocity dispersion  the filtered turbulent velocity dispersion within the $\Lambda=300 \rm ~kpc$ scale,  as a function of the X-ray emission of cells.} 
 \label{fig:scatter0}
\end{figure}

\section{Methods}
\label{sec:methods}

We analyse a high-resolution cluster simulation obtained with the Eulerian code ENZO \citep[][]{enzo14}, already used in previous work \citep[e.g.][]{va18mhd,dom19,bonafede21}. 
This ideal MHD simulation used up to eight levels of Adaptive Mesh Refinement (AMR) to reach a peak spatial resolution of  $\Delta x = 3.95 \rm ~ kpc/cell$ in most of the cluster core, and a resolution of $\Delta x= 15.8 \rm ~ kpc/cell$ in the entire virial volume \citep[][]{va18mhd}.
 We selected this particular cluster simulation because its final mass is close to the one of the real Coma ($M_{200} \approx 1.1\cdot 10^{15} M_{\odot}$ at $z=0.02$). This system has thermal properties that are very similar to the real Coma in terms of entropy and density profiles (see e.g. \citealt[][]{va10kp} and also Sec.\ref{A0}).
This cluster accreted $\sim 50\%$ of its final virial mass during a major merger (i.e. mass ratio larger than $1/5$ of the cluster mass at that time) at around $z \sim 0.45$ \citep[][]{va10kp}, and it has ongoing accretion from smaller clumps at $z=0$, which induce gas perturbation and peripheral shocks, like in the real Coma \citep[e.g.][]{2023A&A...670A.156C}.
We studied this system already to compare with observations of non-thermal properties of the Coma cluster, showing that the simulation produces a Faraday Rotation profile consistent with that measured in the Coma cluster \citep[][]{va18mhd} and that the turbulence generated in the simulation allows reproducing the observed radio halo under the assumption of second-order acceleration of relativistic electrons \citep[][]{bonafede21}.

These radio observables are tightly linked with the level of turbulent gas motions naturally developed by major and minor merger events, which we estimated to contribute to $\sim 5\%$ of the total gas pressure in the form of isotropic turbulent motions in the cluster centre  \citep[][]{va18mhd}.

To compare with recent X-ray observations by the XRISM collaboration, we computed the continuum and line X-ray emission from all cells in the simulation within the  $[5,7] \rm ~keV$ energy rang, assuming  a constant composition and metallicity ($30 \%$ of the solar one) for all cells, with the B-APEC emission model (\url{https://heasarc.gsfc.nasa.gov/xanadu/xspec/manual/Models.html}). For our analysis, we also had to identify the turbulent part of the 3-dimensional gas velocity field in the ICM, with a 
small-scale filtering approach, based on previous work \citep{2006MNRAS.369L..14V,va09turbo,va11turbo,va12filter,va17turb,2020MNRAS.495..864A,2022A&A...658A.149S}.  More sophisticated approaches using a variable filtering length, combined with the masking of velocity jumps induced by shocks, can be used for a more accurate removal of bulk from turbulent motions \citep[][]{va12filter,2020MNRAS.495..864A,2019ApJ...874...42V}. 
However, we have also showed that the result of such more complex filtering is equivalent to a fixed spatial filtering length with a length $ \Lambda_t \sim R_{100}/10$, i.e. $\Lambda_t \sim 300 \rm ~kpc$ for clusters in the  $\sim 10^{15} \rm M_{\odot}$ mass range \citep[see e.g.][Fig.6]{va12filter}, as bulk motions start to contribute significantly to the gas kinetic energy budget in simulated clusters above this scale
\citep[e.g.][]{va11turbo,va17turb,2020MNRAS.495..864A}.

In the remainder of this paper and in line with previous work we estimate the gas turbulent velocity after filtering out motions on scales $\geq \Lambda_t$:   $\delta \vec{v} = \vec{v} - \langle \vec{v} \rangle_{\Lambda_{t}}$,  where $\vec{v}$ is  the local velocity and density and $\langle \vec{v} \rangle_{\Lambda_{t}}$ is the velocity averaged on the $\Lambda_t$ scale.  For completeness, we also present results for different fixed filtering lengths: $\Lambda_t=150, 300, 600$ and $900$ $\rm kpc$.
In order to avoid any confusion, in the remainder of the paper we will explicitly state when these "filtered" turbulent velocities are used, while all X-ray weighted quantities (e.g. the velocity dispersion along the line of sight) will not include any spatial filtering at all.

\section{Results}
\label{sec:res}

We give in Figure \ref{fig:map0} the maps of projected X-ray surface brightness obtained in the [5-7] keV band, of the X-ray weighted mean gas temperature, and of the total velocity and velocity dispersion along the line of sight. 
The latter two are computed in two ways: a) in the top panels, by directly computing the average (volume-weighted) velocity and velocity dispersion along each the LOS, considering a field of view $3.95 \times 3.95 ~\rm kpc^2$ for each LOS (as in the original resolution of the simulation), or b) by computing the X-ray emission-weighted average velocity and velocity dispersion along the LOS after resampling on  $90 \times 90 \rm ~kpc^2$ pixels (corresponding to a XRISM FOV at the distance of the Coma cluster), similar to \citet{Coma_XRISM}. 
The Figure shows the significant difference between the projection of volume-weighted and X-ray-weighted quantities, with the X-ray-weighted quantities generally biased low, due to the combination of density-squared  (X-ray emissivity) weighting, and due to the patchy nature of the turbulence and bulk motions in the simulation.
 \citet{Coma_XRISM} recently concluded that the turbulent spectrum in the Coma cluster may be much steeper than the Kolmogorov model of turbulence. This conclusion is motivated by the fact that a Kolmogorov model fitting their measured velocity structure function\footnote{It shall be noticed that the $\sim \rm Mpc$ scale point in the VSF of  \citet{Coma_XRISM} has been obtained based on the measured velocity offset between galaxies and the gas distribution, which might be partly contributed by bulk motions.}(VSF) would predict a line more than twice wider than the measured one. A spectral distribution of velocity fluctuations overall described by the classical $\propto k^{-5/3}$ (where $k=2\pi/l$ and $l$ is a spatial scale) scaling of Kolmogorov turbulence is instead generally found in all modern  cosmological simulations \citep[][]{do05,va09turbo,va11turbo,va17turb,2022A&A...658A.149S}.
 
The VSF is defined as  $VSF(\vec{r})=\langle |\vec{v_z}({\vec \chi)}-\vec{v_z}(\vec{\chi}+\vec{r})|^2\rangle $, where $\vec{v_z}$ is the velocity along the line of sight and $\vec{r}$ is a displacement with respect to the $\vec{\chi}$ position. The Kolmogorov model predicts a $VSF(r) \propto r^{2/3}$ scaling for isotropic and volume-filling turbulence. 
The intrepretation by \citet{Coma_XRISM} is based on the assumption of a homogeneous turbulent model cluster-volume filling, normalised to the VSF, that is used in combination with the beta-model thermal-density distribution of the Coma cluster to calculate the line width integrated along the LOS (e.g. see also \citealt{2025arXiv251021918E}).
Our simulation allows us to step beyond this simplified model and to compare XRISM observations with the properties of turbulence in our simulated Coma-like cluster.

Figure \ref{fig:VSF} gives the 3-dimensional VSF for the velocity component along the LOS, computed from $10^7$ cells in the cluster volume. In line with previous work \citep[e.g.][]{va11turbo,va17turb,2022A&A...658A.149S},  for more than one order of magnitude in spatial scales the measured VSF is in line with the Kolmogorov model ($\propto r^{2/3}$) \footnote{The gas velocity power spectra in clusters of galaxies simulated in cosmology is known  to have a steepening,  compared to the $\propto k^{-5/3}$, for $\leq 100 \rm ~kpc$ scales. This is partially due to numerical effects: velocity fluctuations on scales smaller than $\leq 8$ grid cells are increasingly damped by numerical dissipation \citep[e.g.][]{pw94,krit11,miniati14}. Moreover, AMR additionally reduces the amount of  small-scale velocity fluctuations, when spectra or VSFs are computed over a large volumes, owing to the mixing of different  AMR levels \citep[e.g.][]{va11turbo}.  Finally, global power spectra of gas motions across the cluster volume are steeper than the ones measured in smaller sub-volumes (which are closer to $\propto k^{-5/3}$), as global spectra mix different turbulent cascades with different normalisations \citep[e.g.][]{2022A&A...658A.149S}.}. To mimic the recent XRISM analysis of Coma, we produced 50 random variations of the 2D estimates of the VSF, based on the X-ray weighted velocity map with $90 \times 90 \rm ~kpc^2$ pixels (as in the last column of Fig.\ref{fig:map0}). In detail, we used triplets of random pixels with a relative projected distance as in the real observation, and we built their VSF by computing the velocity difference between them. The results are shown by the multitude of thin coloured lines (some of which producing even inverted trends of the VSF with increasing scale) in Fig.\ref{fig:VSF}, which are over-imposed to the actual XRISM data point (grey points), to the two models with different slopes discussed in \citet{Coma_XRISM} and with the region of uncertainties of the VSF discussed there.
 Within the large scattered distribution of randomly measured VSFs, a large fraction of them are similar to the recent XRISM measurement, and within their estimated statistical uncertainty and cosmic variance. 
We conclude that our simulated Coma-like cluster is characterized by a 3-dimensional velocity field which is compatible with a Kolmogorov turbulence in a stratified medium across a fairly large range of scales, and that its observable VSFs are in principle compatible with the measurements recently made with XRISM.

Next, we show that the same simulated velocity field also appears capable to reproduce the iron line emission measured by XRISM. In detail,  we simulated the iron line emission from four lines randomly drawn within a $\leq 200 \rm ~kpc$ region from the cluster centre (whose position is shown by the black squares in Fig.\ref{fig:map0}), as shown in Fig.\ref{fig:line}. The line emission profile is generated by adding the X-ray emission in the [5-7]keV band for each cell along the line of sight (assuming a constant metallicity) and by including the $\Delta E/E = v_z/c$ Doppler shift of the line centre, based on the measured velocity along the LOS (again for $90 \times 90 \rm ~kpc^2$ FOV). The additional shaded areas show the best models produced to fit the North-West quadrant of the XRISM observation of Coma, which plausibly includes a bulk motion along the line of sight. Our simulated FOVs generate line shapes with a range of profiles, including cases which closely match the Iron line profile measured by XRISM on Coma (like line 3 for the $z=0.0217$ component in the Coma observation, or line 2 for the additional velocity component detected by XRISM along the LOS), while other have different shapes (e.g. lines 1 and 4).   

This might seem at odds with the conclusions in \citet{Coma_XRISM}, where a Kolmogorov model matching the observed VSF results in a line width that is too large compared to observations. 
The discrepancy arises from the fact that the turbulence in simulations is very different from a homogeneous and cluster-volume filling model assumed in \citet{Coma_XRISM}. Turbulence in galaxy clusters is driven by several accretion and merging events on different spatial- and time-scales. This generates a number of active turbulent regions with different turbulent intensities that might also mix due to turbulent diffusion and advection. This situation generates an intermittent, multiscale and spatially patchy distribution of turbulence with a Kolmogorov-like spectrum across a fairly large range of scales \citep[e.g.][]{do05,in08,va09turbo,va11turbo,miniati14}. As a net result, we find that the effective line width in this situation is reduced by a factor $\sim 2$ compared to that estimated from a homogeneous and cluster-volume filling model with the same turbulent energy budget and spectrum (Sec. \ref{A1}).

 We also remark that this result appears in line with the test shown in the Appendix of 
\citet{Coma_XRISM}, where the emission-weighted velocity dispersion for 14 Coma-like clusters is measured.
In light of these results, another important point is to evaluate the fairness of the turbulence energy budget estimated from the integrated line-width along the LOS. Given the complex properties of turbulence in the simulated clusters, it is not immediately clear which fluctuation scales the XRISM measurements are sensitive to, and what biases arise from emissivity weighting in the inhomogeneous and stratified ICM.
The top panel of Figure \ref{fig:scatter0} gives the distribution of the X-ray weighted velocity dispersion (unfiltered) and of the distribution of "true" turbulence along the line of sight (which is based on our fiducial 3-dimensional small-scale filtering procedure for $\Lambda_t=300 ~\rm kpc$) while the central panel gives the ratio between the two within each $90 \times 90 ~\rm kpc^2$ lines of sight.
In both cases we consider only central ($\leq 500 \rm ~kpc$ from the cluster centre) regions.  While the distribution of the X-ray velocity dispersion is well within the estimates recently produced by \citet{Coma_XRISM}, the distribution of our best estimate of the true (filtered) turbulent content in the same cluster is higher on average, typically by $\sim 40-50 \%$ (at a $\sim 200-300 ~\rm kpc$ radius). The lower panel gives the ratio between the X-ray weighted and the turbulent velocity dispersion as a function of the X-ray emission within FOV: the ratio 
on average decreases as a function of increasing X-ray surface brightness albeit with  a significant scatter which does not allow to extract a simple correction factor,  because the mismatch does not only depend on the X-ray surface brightness,  but rather on the specific gas structures along the LOS.

\begin{figure}
\includegraphics[width=0.495\textwidth]{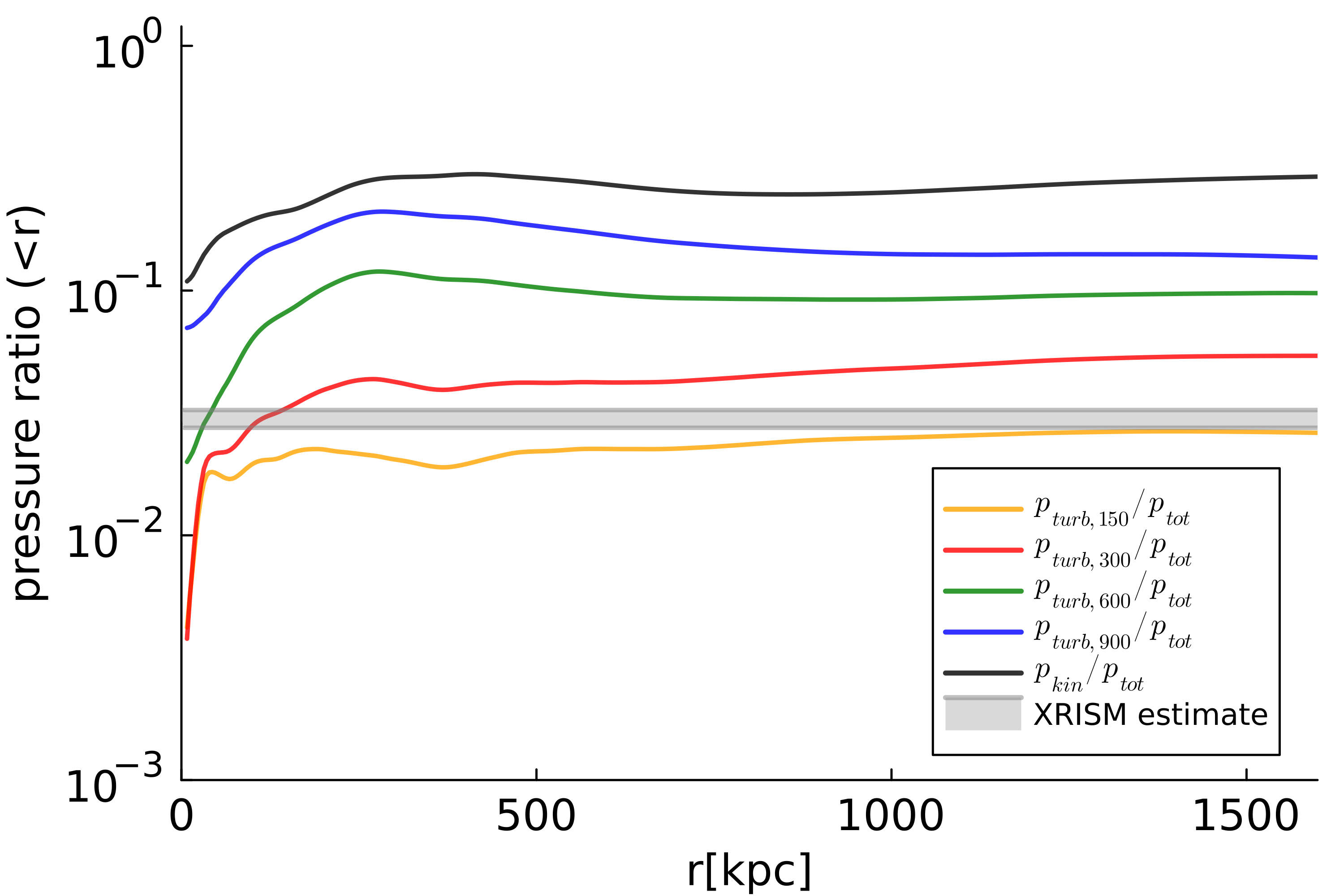}
\caption{3-dimensional radial profile of the non-thermal to total pressure ratio within the radius, considering  the total (unfiltered) gas velocity field ($p_{kin}$) or four different choices for the small-scale filtering length ($\Lambda=150, 300, 600$ and $900$ $\rm kpc$, where $p_{turb,300}$ is our reference estimate as discussed in the main text). The  horizontal grey strip gives the estimate for the Coma cluster by \citet{Coma_XRISM}.}

 \label{fig:scatter1}
\end{figure}

Finally, we re-assess the amount of the non-thermal pressure support in this cluster, to closely compare with the result by \citet{Coma_XRISM}, who reported a  kinetic pressure support $\approx 3\%$ of the total pressure in the cluster centre. In  Figure \ref{fig:scatter1} we give the 3-dimensional profile of the ratio between turbulent and total pressure within the radius.
 We use the same formula by \citet{Coma_XRISM}:  $p_{turb,kin}/p_{tot}=[1+3/(\gamma \mathcal{M}_{3D}^2)]^{-1}$ where $\mathcal{M}_{3D}$ is the local 3-dimensional Mach number measured by dividing either the turbulent velocity dispersion  or the total velocity dispersion for the sound speed of each cell.  Here we show the estimates of the non-thermal pressure support by turbulent motions obtained using four different filtering lengths if increasing size, i.e. from $\Lambda_t=150 ~\rm kpc$ to $=900 ~\rm kpc$ (with $\Lambda_t=300 ~\rm kpc$ being our fiducial choice due to the large contamination by bulk motions on larger scales, as discussed in Sec.~\ref{sec:methods}).
  In all cases the non-thermal pressure ratio increases with radius, as expected \citep[][]{va18turb}, with the total (unfiltered) kinetic pressure reaching $\sim 30 \%$ at large radii, and with the turbulent kinetic pressure ranging from $\sim 3 \%$ to $\sim 20\%$, depending on the adopted filtering length. For our fiducial estimate of turbulence in clusters of this size ($\Lambda_t=300 ~\rm kpc$) we get $\sim 5\%$ of non-thermal pressure within $1.5 \rm ~Mpc$ from the cluster centre, which is $\approx R_{500}$ for this cluster.
  This budget is 
  higher than what is estimated for the Coma cluster by \citet{Coma_XRISM}, but is also line with our previous estimates for this, and similar simulated clusters \citep[e.g.][]{va11turbo,va18mhd}. 
 Based on the relative distribution between X-ray weighted and true turbulence along the line of sight, we conclude that actual turbulent non-thermal support in the ICM is significantly higher (typically by a factor $\sim 2$) than the  estimate given by spectroscopic X-ray analysis, which is in turn biased low by the smaller range of scales sampled through X-ray emission. 

As a final remark, we note that turbulent density fluctuations in the ICM might be constrained also from projected X-ray surface brightness fluctuations  \citep[e.g.][]{2014A&A...569A..67G, 2014ApJ...788L..13Z, 2017ApJ...843L..29E,2020MNRAS.493.5838M,2021MNRAS.500.5072M}, although the relation between velocity and density fluctuations is more complex in realistic ICM conditions (\citealt{2022A&A...658A.149S}, see also the discussion in \citealt[][]{Coma_XRISM}). A combined analysis of X-ray surface brightness fluctuations and line broadening generated by our simulations has the potential to further shed light on the properties of turbulence and will be studied with future work.

\section{Conclusions}
\label{sec:conclusions}

The physics of turbulence in the weakly collisional ICM is expected to be more complicated than that in fluid numerical simulations, where an idealised collisional plasma is assumed.  In this article we have investigated whether the recently observed properties of turbulent gas motions inferred via X-ray spectroscopy by \citet{Coma_XRISM} are alrady in tension with the idealised view from numerical simulations on large sales ($\geq 10 \rm ~kpc$) and have already started to pinpoint aspects driven by the collisionless nature of the ICM, which are expected to arise on smaller scales \citep[e.g.][]{2011MNRAS.410.2446K,bl11,2014ApJ...781...84S,bj14,2016PNAS..113.3950R,2018ApJ...863L..25S,2020JPlPh..86e9003S,2025A&A...694A..25K}.

In particular, using a homogeneous and cluster-volume filling model of turbulence, 
the VSF and X-ray line width measured by XRISM require a turbulent spectrum that is much steeper than a Kolmogorov, possibly pin-pointing either a very large effective viscosity in the ICM or a situation where large scale motions generated by recent mergers have not yet cascaded down to small scales. Using a high-resolution cosmological simulation of a Coma-like cluster  (Sec.\ref{sec:methods}), we have shown that :
\begin{itemize}
\item  the spatially patchy, not homogeneous  and multi-scale turbulence in the simulation generates a VSF and X-ray line-widths integrated along the LOS that are both consistent with the XRISM observations, thus releasing the request of a turbulent spectrum that is much steeper than a Kolmogorov. This also shows that the interpretation of XRISM data is very sensitive to the assumed turbulent model.
\item  X-ray lines integrated along the LOS are sensitive to the X-ray brightest regions. This biases the effect of line broadening due to turbulence to an extent that depends on the properties of the turbulence itself in the integrated volume. In our simulation we find that the simulated widths of XRISM X-ray lines are biased low by $\sim 50\%$ with respect to the turbulent velocity dispersion on scale $\sim 300 \rm ~kpc$ that is measured along the LOS. This bias arises from two factors : the limited range of scales of the turbulent spectrum sampled in this way, and the limited radial extent from the cluster centre which can be probed by X-ray observations (which are biased towards the innermost dense cluster regions). 

\item The pressure support in our simulated cluster from turbulence on filtered scales of 300-600 kpc is estimated in the range $\sim 5-10\%$, which is $\sim 2-3$ times larger than that estimated from the widths of X-ray lines obtained from the simulation,  which are consistent with those measured by \citet{Coma_XRISM}. The actual turbulent pressure support in the ICM can plausibly higher than what is estimated from X-ray spectroscopy, although it is not possible to derive a simple correction factor, due to the large measured scatter in the relation between the true and the X-ray weighted turbulent dispersion along the LOS \citep[see also, e.g.][for valuable numerical work on this issue]{2018A&A...618A..39R,2022A&A...663A..76B,2025arXiv250614441L}.
\end{itemize}

 In summary, we suggest that the fluid turbulence typically produced by cosmological simulations of galaxy clusters (including our ones and the majority of those in the literature) is compatible with the latest XRISM measurements in Coma, and possibly in other clusters, provided that the realistic degree of non-homogeneity of the ICM, the multi-scale nature of the ICM turbulent flows and the biases inherent to X-ray spectroscopy are fully  taken into account. 
 
 More in general, our results highlights the importance of advanced numerical simulations to support the interpretation of the data from X-ray calorimeters in the future.

\section*{acknowledgements}
FV has been partially supported by Fondazione Cariplo and Fondazione CDP, through grant n° Rif: 2022-2088 CUP J33C22004310003 for the "BREAKTHRU" project. GB acknowledges partial support from INAF Theory Grant "Theory and simulations of non-thermal phenomena in galaxy clusters and beyond". F.V. acknowledges the CINECA award  "IscrB\_CREW"  under the ISCRA initiative, for the availability of high-performance computing resources and support. We wish to thank our reviewer for the valuable comments on our draft, and S. Ettori, D. Eckert, T. Lebeau, A. Ignesti, F. Rincon,  M. Markevitch, V. Biffi, J. Zuhone and I. Zhuravleva for helpful comments and discussons. The authors wish to declare that $0\%$ of their original manuscript has been produced using AI of any sort.

\bibliographystyle{aa}
\bibliography{franco3}

\appendix
\section{Comparison with radial profiles of the Coma cluster}
\label{A0}

\begin{figure}
\includegraphics[width=0.495\textwidth]{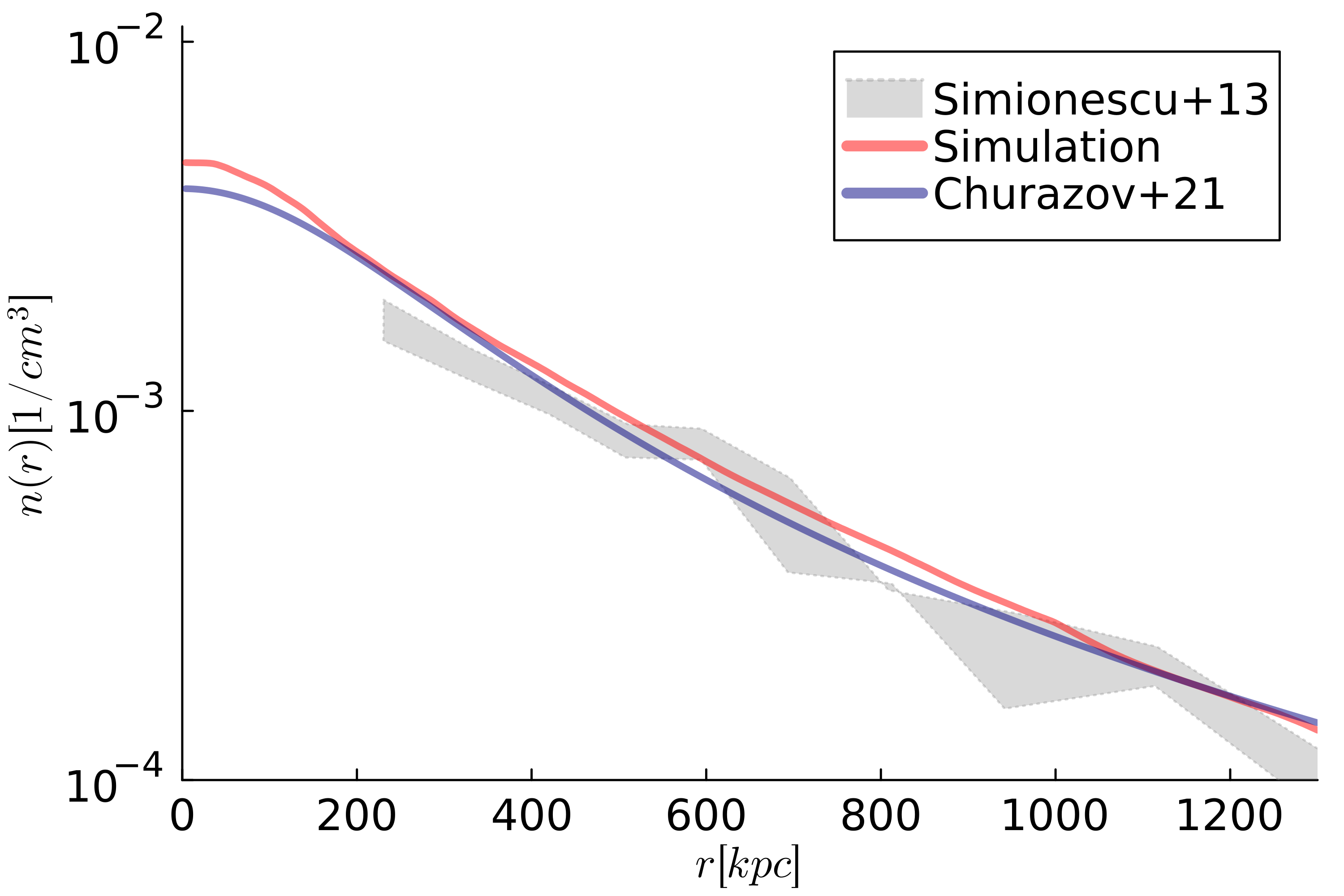}
\includegraphics[width=0.495\textwidth]{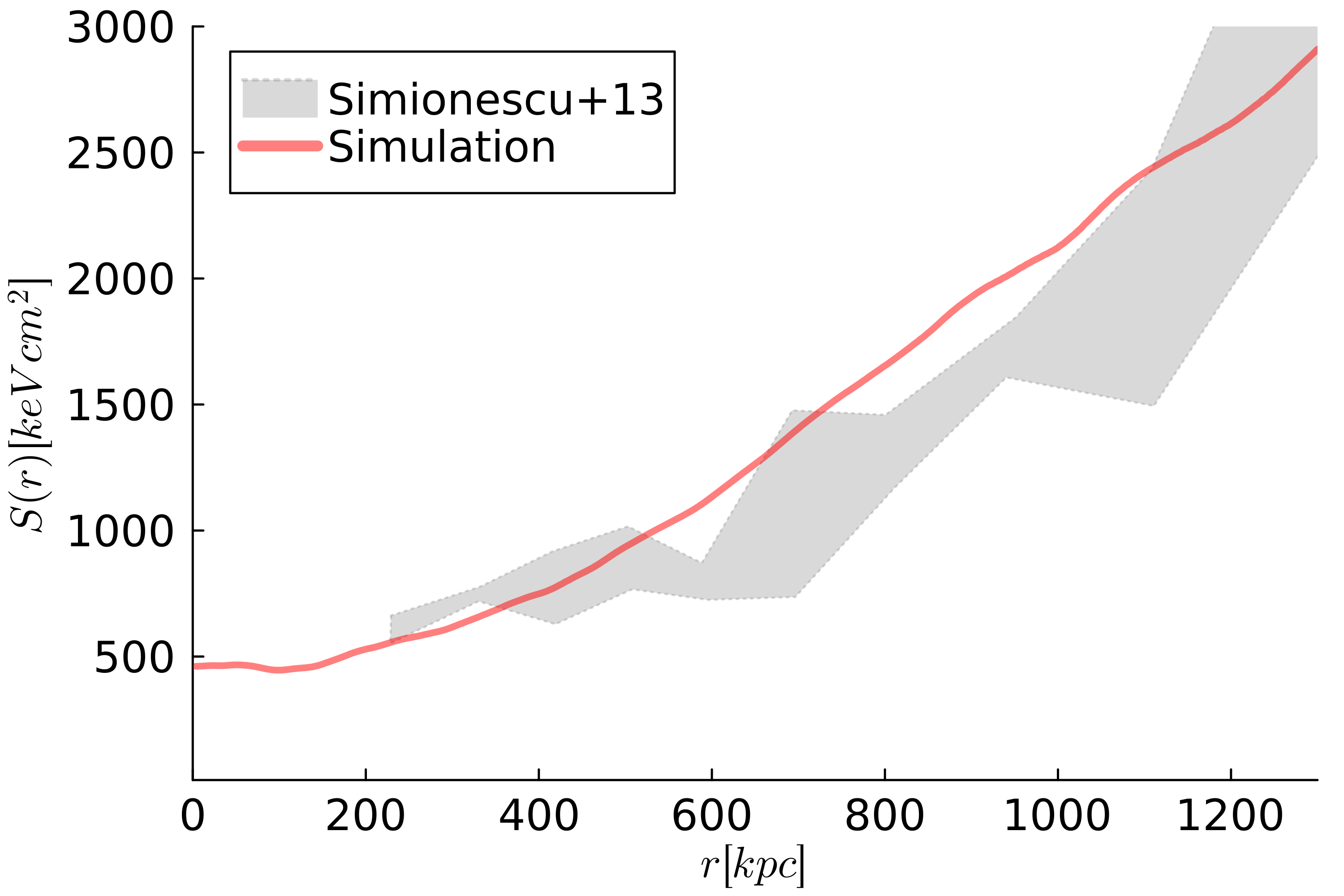}
\caption{Average radial density (top) and gas entropy (bottom) profile for our simulated  cluster at $z=0.02$, compared with recent observational constraints for the real Coma cluster derived by \citet{simionescu13} using SUZAKU  (the grey areas contain the two measurements taken within two narrow sectors in the outskirts of Coma), and by \citet{churazov21} using eROSITA.}
 \label{fig:A0}

\end{figure}

We compared the radial profiles of gas density, $n(r)$ and gas entropy (defined as $S(r)=T(r)/n(r)^{2/3}$ where $T$ is the gas temperature) from our Coma-like simulated cluster,   with the available observational constraints for the real Coma. This is shown by the two panels in Fig.\ref{fig:A0}. The simulated density profile compares reasonably well with $\beta$-model parametrisation recently obtained by \citet{churazov21} using X-ray observations by the eROSITA satellite, and also with the pointed deep X-ray exposures of two narrow selections along the E+NE and NW+W directions in the Coma cluster, take by \citet{simionescu13} using SUZAKU. The lower panel shows that the average gas entropy profile of our simulated cluster is also fairly in line with the most recent constrains obtained by \citet{simionescu13} using SUZAKU. 
The reasonably good comparison with these two constraints for the real Coma cluster is reassuring as it shows that the gas matter distribution (and its related X-ray emission) from our model is realistic, and also that the entropy stratification of our simulated cluster (which in turns is related to buoyancy and gas mixing as a function of radius) is in line with observations.

\section{Comparison with idealised turbulence}
\label{A1}
\begin{figure}
\includegraphics[width=0.495\textwidth]{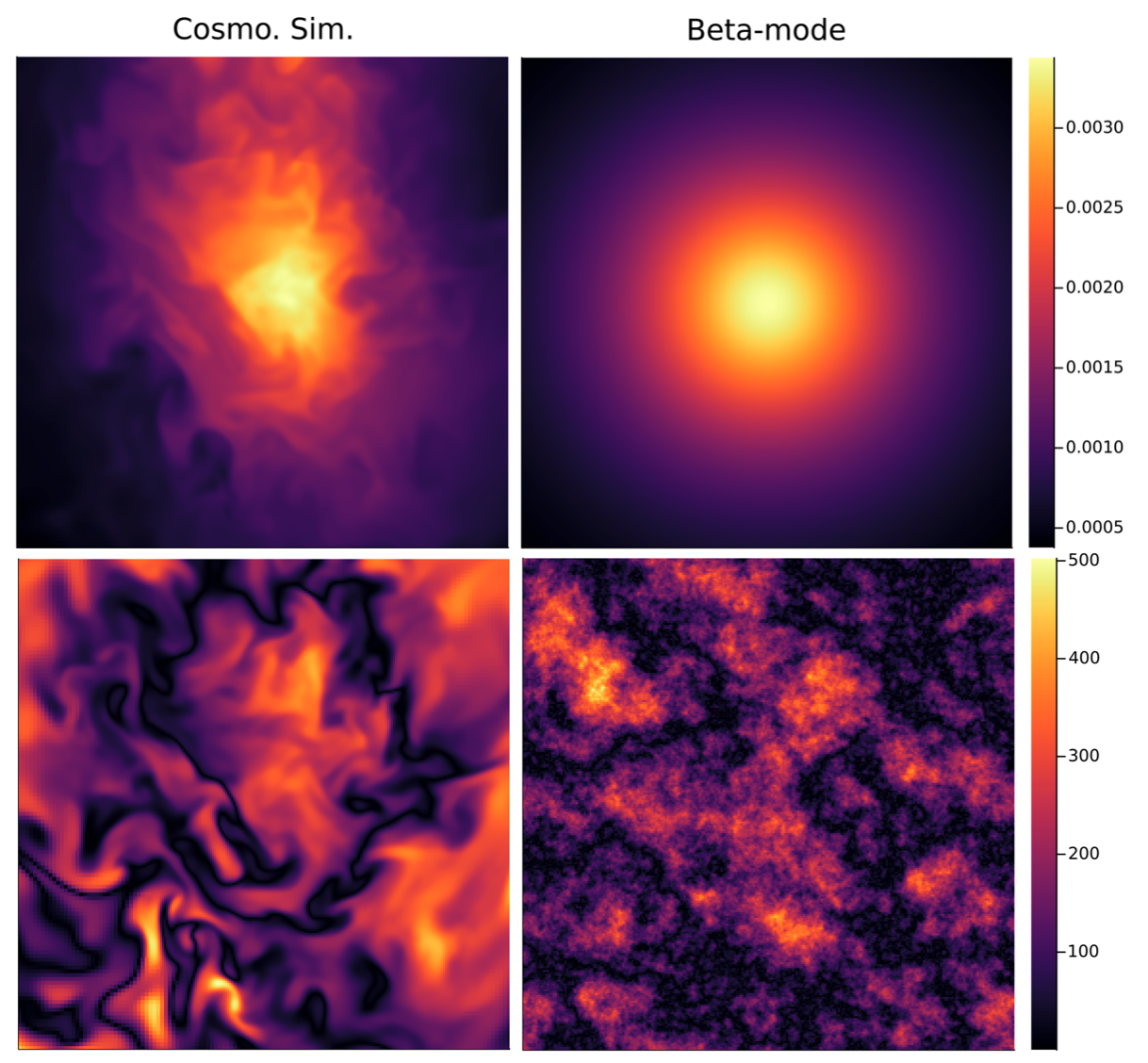}
\caption{Top panels: central slice showing the distribution of gas density (in units of [$\rm part/cm^3$]) for the clusters simulation analysed in the main article, and for a $\beta$-model of the Coma cluster instead. Lower panels: distribution of the amplitude of the gas velocity field (in units of [$\rm km/s$] for the same cosmological simulation (left) or for a 3-dimensional velocity field generated in Fourier space so that it follows the Kolmogorov spectrum for homogenous and stationary turbulence. Each panel has a side $1.6 \rm ~Mpc$. }
 \label{fig:A1}
\end{figure}

\begin{figure}
\includegraphics[width=0.495\textwidth]{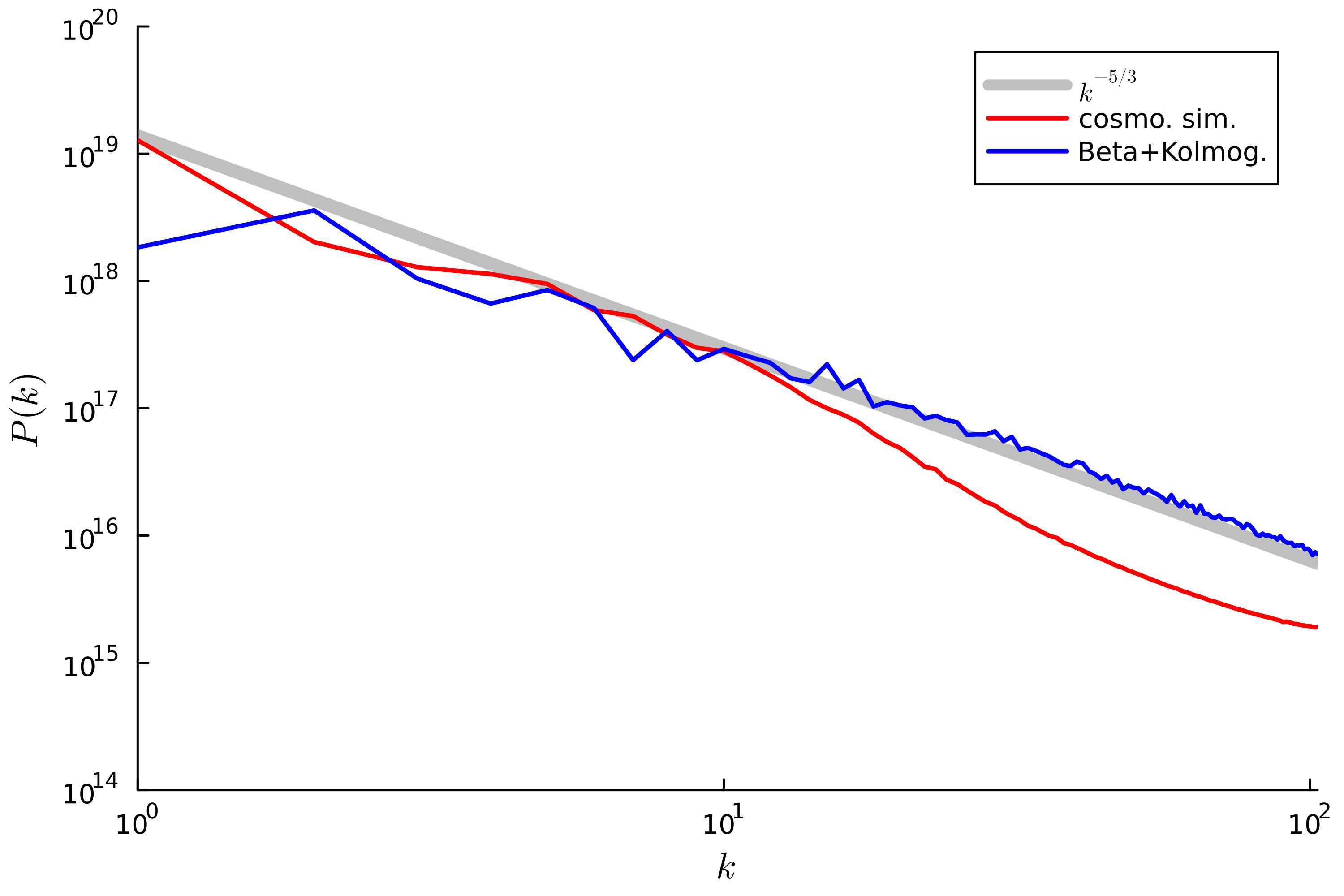}
\caption{Power spectrum (in arbitrary units) of the velocity field for the two cases discussed in Sec.\ref{A1}, with the additional $\propto k^{-5/3}$ shown for comparison.}
 \label{fig:A2}
\end{figure}

In order to quantify the importance of the turbulent model, here we present the comparison between the line width obtained by integrating the simulation along the LOS and that obtained using a homogeneous and volume filling Kolmogorov turbulent model, similar to the procedure adopted in \citet{Coma_XRISM} and in \citet{2025arXiv251021918E}. 
To calculate the homogenous model, we initialised a spherically symmetric $\beta$-model gas density profiles with parameters of the Coma cluster \citep[e.g.][and references therein]{churazov21} for a $1.6^3  \rm Mpc^3$ volume using $400^3$ cells, and we generated a random 3-dimensional velocity field in  Fourier domain, assuming it follows the Kolmogorov scaling ($P(k) \propto k^{-5/3}$) from the largest to the smallest available grid scales. In order to obtain a random Gaussian distribution of velocities, we randomly extracted vectors components from the Rayleigh distribution. 
The 3-dimensional velocity field within the $1.6^3  \rm Mpc^3$ volume  generated in such a way is renormalised to exactly match the kinetic energy budget, within an equivalent volume centred on the cluster centre, in our cosmological simulation.
Figure \ref{fig:A1} shows the distribution of gas density and velocity along the line of sight for two middle planes in the two datasets, while Fig. \ref{fig:A2} shows the comparison of the 3-dimensional velocity power spectra of the two datasets.  The latter confirms the trend of the VSF of Fig.\ref{fig:VSF}: the spectrum of gas motions in the simulated cluster approximately follows the Kolmogorov spectrum up to $\sim 1.6 \rm ~Mpc$, while it steepens for $\leq 100 \rm ~kpc$ scales (see Sec.\ref{sec:res} for a discussion).

\begin{figure}
\includegraphics[width=0.495\textwidth]{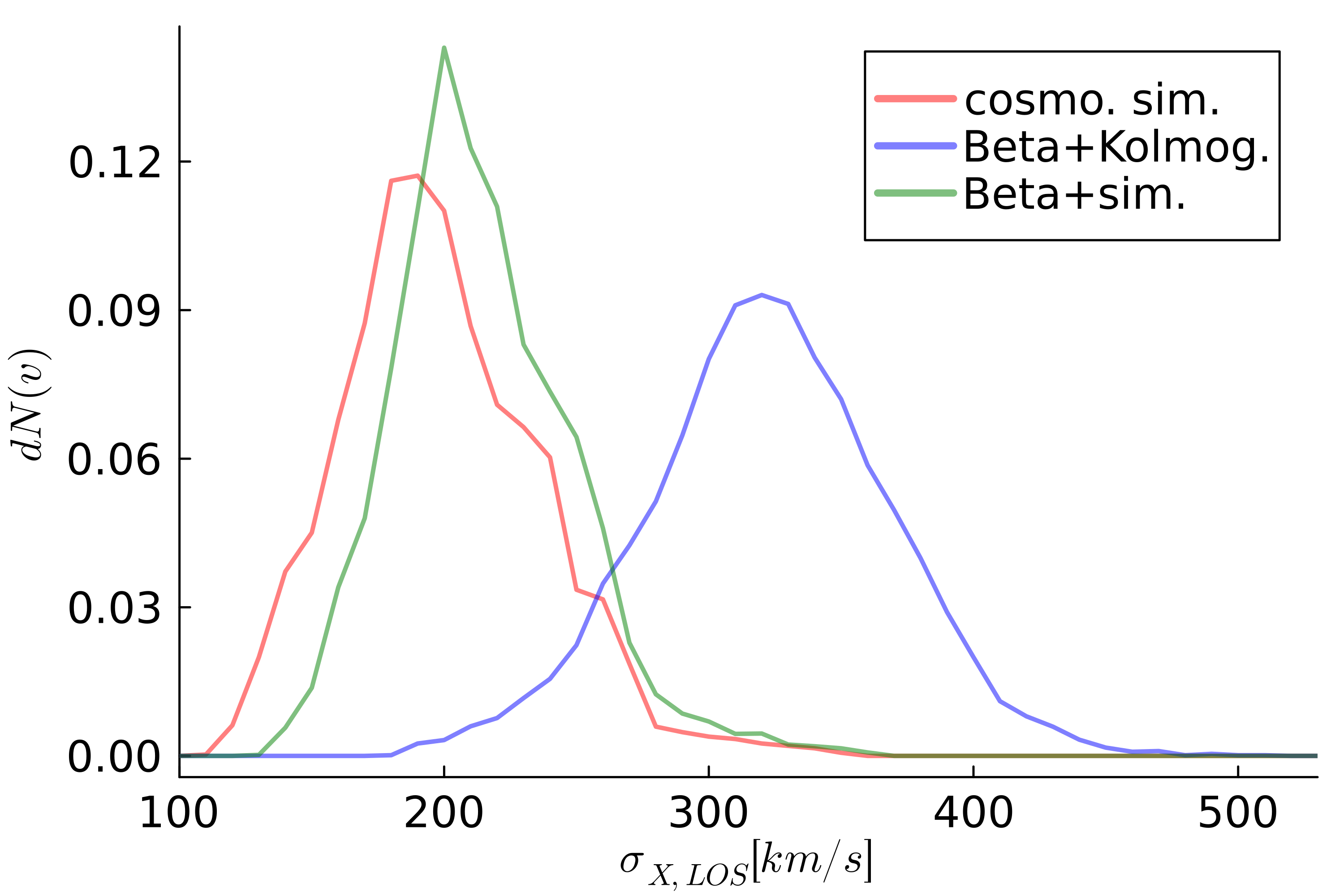}
\caption{Comparison of the distribution of the X-ray weighted gas velocity dispersion along $10^4$ simulated lines of sight through the three datasets considered in Sec.\ref{A1}: a symmetric $\beta$-model distribution of density combined with a homogenous random velocity field (blue), the density and velocity field taken from our cluster simulation (red), and an additional "mixed" model which combines the density distribution of the first model with the velocity field of the second model.}
 \label{fig:A3}
\end{figure}

In Figure \ref{fig:A3} we show the distribution of the X-ray weighted gas velocity dispersion along $10^4$ random lines of sight through the two simulated datasets. The distribution derived from the cosmological simulation of our cluster is clearly peaked at a smaller value, $\sigma_{X,LOS}\sim 180 \rm ~km/s $, while the distribution derived from the idealised turbulent model peaks at the higher value of $\sigma_{X,LOS}\sim 330 \rm ~km/s $. 

This simple experiment (although limited to the central fraction of the simulated volume, and not to the full depth of the lines of sight used in  Fig.\ref{fig:scatter0}) very well illustrates that, for a realistic dynamical model of turbulence in the ICM, a smaller typical broadening of X-ray lines is to be expected, compared to a more idealised model of turbulence, despite the total kinetic energy of the two models is the same. The key factor here is the realistic level of intermittency of turbulence naturally produced by the multi-scale accretions onto cluster simulated in cosmology: turbulence has in this case a smaller filling factor compared to fully homogenous models which assume a large injection scale and stationary driving.  For the same total kinetic energy budget, a more intermittent turbulence distribution produces on average a smaller velocity dispersion along the line of sight.  Conversely, if two models produce on average the same  broadening of X-ray lines along the line of sight, homogenous model is biased to be characterised by significantly smaller kinetic energy budget. Finally, with an additional test (green line) we show the distribution of X-ray weighted gas velocity dispersions obtained in a "mixed" model, in which we use the $\beta$-model density distribution to compute the X-ray emission, and combine this with the 3-dimensional velocity field of the cosmological simulation. The distribution obtained in this way is quite close the one obtained using the density distribution from the simulation, and it shows that the key factor here is the intermittency in the turbulent velocity field, and not in the density field.

\end{document}